\documentclass[
    ,final            
    ,sort&compress    
  ]
  {aipproc}

\layoutstyle{6x9}

\begin{document}

\title{Effects of constraints in general branched molecules: A quantitative ab initio study in HCO-L-Ala-NH$_2$}


\author{Pablo Echenique}{
  address={Departamento de F\'{\i}sica
           Te\'orica, Facultad de Ciencias, Universidad de Zaragoza,
           \\ Pedro Cerbuna 12, 50009, Zaragoza, Spain.}
  ,altaddress={Instituto de Biocomputaci\'on y
               F\'{\i}sica de los Sistemas Complejos (BIFI), \\ Edificio
               Cervantes, Corona de Arag\'on 42, 50009, Zaragoza, Spain.}
}

\author{J. L. Alonso}{
  address={Departamento de F\'{\i}sica
           Te\'orica, Facultad de Ciencias, Universidad de Zaragoza,
           \\ Pedro Cerbuna 12, 50009, Zaragoza, Spain.}
  ,altaddress={Instituto de Biocomputaci\'on y
               F\'{\i}sica de los Sistemas Complejos (BIFI), \\ Edificio
               Cervantes, Corona de Arag\'on 42, 50009, Zaragoza, Spain.}
}

\author{Iv\'an Calvo}{
  address={Departamento de F\'{\i}sica
           Te\'orica, Facultad de Ciencias, Universidad de Zaragoza,
           \\ Pedro Cerbuna 12, 50009, Zaragoza, Spain.}
  ,altaddress={Instituto de Biocomputaci\'on y
               F\'{\i}sica de los Sistemas Complejos (BIFI), \\ Edificio
               Cervantes, Corona de Arag\'on 42, 50009, Zaragoza, Spain.}
}

\begin{abstract}
A general approach to the design of accurate classical potentials for
protein folding is described. It includes the introduction of a
meaningful statistical measure of the differences between
approximations of the same potential energy, the definition of a set
of Systematic and Approximately Separable and Modular Internal
Coordinates (SASMIC), much convenient for the simulation of general
branched molecules, and the imposition of constraints on the most
rapidly oscillating degrees of freedom. All these tools are used to
study the effects of constraints in the Conformational Equilibrium
Distribution (CED) of the model dipeptide HCO-L-Ala-NH$_2$. We use ab
initio Quantum Mechanics calculations including electron correlation
at the MP2 level to describe the system, and we measure the
conformational dependence of the correcting terms to the naive CED
based in the Potential Energy Surface (PES) without any simplifying
assumption. These terms are related to mass-metric tensors
determinants and also occur in the Fixman's compensating potential. We
show that some of the corrections are non-negligible if one is
interested in the whole Ramachandran space. On the other hand, if only
the energetically lower region, containing the principal secondary
structure elements, is assumed to be relevant, then, all correcting
terms may be neglected up to peptides of considerable length. This is
the first time, as far as we know, that the analysis of the
conformational dependence of these correcting terms is performed in a
relevant biomolecule with a realistic potential energy function.
\end{abstract}

\maketitle

\section{Introduction}
\label{sec:introduction}

Proteins are long chains comprised of twenty different amino acidic
monomers and they are central elements in the biological machinery of
all known living beings. They perform most of the catalytic tasks that
are vital in the many coupled chains of chemical reactions occurring
in the cells, they are found as structural building blocks in the
cytoskeleton or in organelles, such as the ribosome, and they also
play a very important role as membrane receptors. Their absence or
malfunctioning is related to many diseases such as Creutzfeldt-Jakob's
or Cancer \cite{PE:Dob2002NAT,PE:Kel2002NSB} and the proteins involved
in the biology of pathogens are often the preferred target of newly
designed drugs (see the talks by E. Freire and C. Cavasotto in this
meeting).

Despite their complexity and the many opposing forces that determine
their behaviour, these molecules swiftly acquire a unique
three-dimensional \emph{native} structure in the physiological
milieu. Some details of this process are still not clear (see the talk
by J. M. S\'anchez-Ruiz), such as the relative proportion of the
naturally occurring proteins that fold co- or post-translationally
(i.e., during or after biosynthesis at the ribosome)
\cite{PE:Bas2003JCMM}, or the role played by molecular chaperones such
as GroEL (see M. Karplus' talk), the Prolyl-peptidyl-isomerase, or the
Protein disulfide isomerase, among others. However, since the
pioneering work of Anfinsen \cite{PE:Anf1973SCI}, it is known that a
large number of water-soluble globular proteins are capable of
reaching their native structure \emph{in vitro} after being unfolded
by changes in their environment, such as a raise of the temperature,
the addition of denaturing agents or a change in the $pH$. It is the
prediction of the native structure in these cases (only from the amino
acid sequence and the laws of physics) that has become paradigmatic
and receives the name of \emph{protein folding problem}.

In 2005, in a special section of the Science magazine entitled ``What
don't we know?'' \cite{PE:What2005SCI}, a selection of the hundred
most interesting yet unanswered scientific questions was
presented. What indicates the importance of the protein folding
problem is not the inclusion of the question \emph{Can we predict how
proteins will fold?}, which was a must, but the large number of other
questions which were related to or even dependent on it, such as
\emph{Why do humans have so few genes?},
\emph{How much can human life span be extended?},
\emph{What is the structure of water?},
\emph{How does a single somatic cell become a whole plant?},
\emph{How many proteins are there in humans?},
\emph{How do proteins find their partners?},
\emph{How do prion diseases work?},
\emph{How will big pictures emerge from a sea of biological data?},
\emph{How far can we push chemical self-assembly?} or
\emph{Is an effective HIV vaccine feasible?}, to quote just a few of them.

Some authors \cite{PE:Dag2003NRMCB} divide the problem in two parts:
the prediction of the three-dimensional, biologically functional,
native state of a protein and the description of the actual folding
process that takes the protein there from the unfolded state. The
first part, which is more pressing and more technologically oriented,
is included in the second part and it is, therefore, easier to tackle,
as the relative success of knowledge-based methods suggests
\cite{PE:Mou2005COSB,PE:Roh2004ME}. However, we believe that, not only
much theoretical insight may be gained from a solution of the more
general second part of the problem, but also much engineering and
design power, as well as new comprehension about so distinct topics as
the ones quoted in the preceding paragraph. This is why our approach
is one of bottom-top and ab initio flavor.

\section{Potential energy functions}
\label{sec:potentials}

The fundamental theory of matter that is nowadays accepted as correct
by the scientific community is Quantum Mechanics. For the study of the
conformational behaviour of a molecule consisting of $n$ atoms, with
atomic numbers $Z_{\alpha}$ and masses $M_{\alpha}$,
$\alpha=1,\ldots,n$, one typically assumes that relativistic effects
are negligible\footnote{Which, in organic molecules, is approximately
correct for all the particles involved, except, maybe, for some core
electrons in the heaviest atoms.} and that, according to the
Born-Oppenheimer scheme \cite{PE:Bor1927APL}, the great differences in
mass between electrons and nuclei allows to consider that the former
are described by a Hamiltonian which is adiabatically decoupled from
the nuclear one and that depends only parametrically on the positions
of the nuclei. Hence, the behaviour of the system \emph{in vacuum} may
be extracted from the non-relativistic time-independent nuclear
Schr\"odinger equation:

\begin{equation}
\label{eq:Schroedinger}
\Bigg(-\sum_{\alpha=1}^{n}\frac{{\hbar}^{2}}{2M_{\alpha}}
       {\nabla}_{\alpha}^{\,2} +
       \sum_{\beta>\alpha}
       \left ( \frac{e^{2}}{4{\pi}{\epsilon}_{0}} \right ) 
       \frac{Z_{\alpha}Z_{\beta}}{|\vec{R}_{\beta}-\vec{R}_{\alpha}|}
       + E_{\mathrm{e}}^{\,0}(R) \Bigg) \Psi_{\mathrm{N}}(R) =
       E \, \Psi_{\mathrm{N}}(R) \ ,
\end{equation}

where $E_{\mathrm{e}}^{\,0}(\vec{R})$ denotes the effective potential
due to the electronic cloud in the fundamental energy
state\footnote{This additional assumption that the electrons are in
the fundamental state prevents us from describing the catalytic
behaviour of most enzymes, however, the only interest here is to
describe the folding process.} and $R$ is shorthand for
$\vec{R}_{1},\ldots,\vec{R}_{n}$.

Despite the exponential growth in computing power that has been taking
place in the last decades (see, for example, the talks by A. Perczel
and I. Campos), a precise description of the behaviour of any
biologically interesting system derived from the solution of
\eqref{eq:Schroedinger} remains far from being even imaginable. Not to
mention the huge complications that arise when the unavoidable
inclusion of solvent is considered. This is why, omitting a myriad of
possible intermediate descriptions, the most popular choice for the
\emph{in silico} prediction of the protein folding process has become
the use of the so-called \emph{force fields}
\cite{PE:Bro1983JCC,PE:Cor1995JACS,PE:Jor1988JACS,PE:Hal1996JCCa},
in which one assumes that the behaviour of the macromolecule (omitting
again the solvent, to compare with \eqref{eq:Schroedinger}) is
\emph{classical} and may be described via a very simple potential
energy function which, typically, has the form

\begin{eqnarray}
\label{eq:Vff}
V_{\mathrm{ff}} & := &
   \frac{1}{2}\sum_{\alpha = 1}^{N_{r}}
      K_{r_{\alpha}}(r_{\alpha}-r_{\alpha}^{0})^{2} + 
   \frac{1}{2}\sum_{\alpha = 1}^{N_{\theta}}
      K_{\theta_{\alpha}}(\theta_{\alpha}-\theta_{\alpha}^{\,0})^{2} 
   + \sum_{\alpha = 1}^{N_{\phi}} A_{\alpha}
   \cos (B_{\alpha}\phi_{\alpha}+\phi_{\alpha}^{0}) + \nonumber \\
 & + & \sum_{\beta>\alpha}
       \left( \frac{C^{\,\alpha\beta}_{12}}{R^{\,12}_{\alpha\beta}} -
       \frac{C^{\,\alpha\beta}_{6}}{R^{\,6}_{\alpha\beta}} \right)
       + \sum_{\beta>\alpha}
       \left ( \frac{e^{2}}{4{\pi}{\epsilon}_{0}} \right ) 
       \frac{Z_{\alpha}Z_{\beta}}{R_{\alpha\beta}} \,
\end{eqnarray}

where $r_{\alpha}$ are bond lengths, $\theta_{\alpha}$ are bond
angles, $\phi_{\alpha}$ are dihedral angles\footnote{For the sake of
simplicity, no harmonic terms have been assumed for out-of-plane
angles or for hard dihedrals, such as the peptide bond angle $\omega$}
and $R_{\alpha\beta}$ denotes the interatomic distances. Finally, all
the parameters entering \eqref{eq:Vff} (which may amount to thousands)
are customarily fitted to reproduce thermodynamical measurements or
taken from quantum mechanical calculations.

While it is true that these empirical potentials may be detailed
enough to deal with simple conformational transitions in already
folded proteins (see the talk by J. Luque) or with collective motions
of systems of many proteins (see M. Karplus' talk), and that they may
also be used as scoring functions for protein design (as in the talk
by A. Jaramillo), all these applications require only that the
energetics of the native structure and its surroundings be correctly
described. As A. Tramontano told us in her talk, the usefulness of
these simple potentials for \emph{de novo} structural prediction
(assessed via the CASP contest\footnote{See \
\url{http://predictioncenter.org}}) remains much limited.

We believe that one of the reasons of this failure is the lack of
accuracy of the potential energy functions used, since, even if the
parameters fit is properly carried out, the choice of the very
particular dependencies, for example those in \eqref{eq:Vff},
constitutes a heavy restriction in the space of
functions. Accordingly, one of our aims is the design of classical
potentials which are as similar as possible to the effective
Born-Oppenheimer one in \eqref{eq:Schroedinger}. To do this, one must
calculate the electronic energy $E_{\mathrm{e}}^{\,0}(R)$ using the
powerful tools of Quantum Chemistry (see the talks by A. Perczel,
J. J. Dannenberg and M. Amzel) and devise numerically efficient
approximations to it.

In any case, in order to walk the long path connecting Quantum
Mechanics and a classical description amenable to nowadays computers,
one must have a meaningful way of comparing different approximations
of the potential energy of a system. Much in the spirit of the talk by
M. Wall, and using the fact that the complex nature of biological
molecules suggests the convenience of statistical analyses, we have
designed in \cite{PE:Alo2006JCC} a \emph{distance}, denoted by
$d_{12}$, between any two different potential energy functions,
$V_{1}$ and $V_{2}$, that, from a working set of conformations,
measures the typical error that one makes in the \emph{energy
differences} if $V_{2}$ is used instead of the more accurate $V_{1}$,
admitting a linear rescaling and a shift in the energy reference.

This distance, which has energy units, presents better properties than
other quantities customarily used to perform these comparisons, such
as the energy RMSD, the average energy error, etc. It may be related
to the Pearson's correlation coefficient by

\begin{equation}
\label{eq:d}
d_{12} = \sqrt{2}\,{\sigma}_{2}(1-r_{12}^{2})^{1/2} \  .
\end{equation}

Finally, due to its physical meaning, it has been argued in
\cite{PE:Alo2006JCC} that, if the distance between two different
approximations of the energy of the same system is less than $RT$, one
may safely substitute one by the other without altering the relevant
dynamical or thermodynamical behaviour.

\section{Effects of constraints}
\label{sec:constraints}

Another reason underlying the difficulties faced in the computational
study of the protein folding problem is that the large number of
degrees of freedom brings up the necessity to sample an astronomically
large conformational space \cite{PE:Dil1999PSC}. In addition, the
typical timescales of the different movements are in a wide range and,
therefore, demandingly small timesteps must be used in Molecular
Dynamics simulations in order to properly account for the fastest
modes \cite{PE:Sch1997ARBBS}, which lie in the femtosecond range;
whereas the folding of a large protein may take seconds. In order to
deal with these problems, one may naturally consider the reduction of
the number of degrees of freedom describing macromolecules via the
imposition of constraints.

To manage this situation, we have made progresses in two
directions. First, we have devised \cite{PE:Ech2006JCCaNOT} a set of
internal coordinates called \emph{SASMIC} (standing for
\emph{Systematic and Approximately Separable and Modular Internal
Coordinates}), which are much convenient to describe branched
molecules and, specially, polypeptides, without having to rewrite the
whole Z-matrix upon addition of new residues to the chain, and also
allow to maximally separate the soft and hard movements\footnote{An
automatic Perl script that generates the SASMIC Z-matrix, in the
format of typical Quantum Chemistry packages, such as GAMESS or
Gaussian03, from the sequence of amino acids, may be found at \
\texttt{http://neptuno.unizar.es/files/public/gen\_sasmic/}}.

Second, we have used these coordinates, the distance discussed before
and the factorization of the external variables in the mass-metric
determinants that we describe in \cite{PE:Ech2006JCCcNOT}, to study
the possibility of neglecting the conformational dependence of the
correcting terms that appear in the equilibrium distribution of
organic molecules \cite{PE:Ech2006JCCbNOT}.

Constraining the hard coordinates $q^{I}$ to be specific functions
$f^{I}(q^{i})$ of the soft ones (which defines a hypersurface $\Sigma$
in the whole conformational space) produces two \emph{classical}
constrained models which are known to be conceptually
\cite{PE:Ral1979JFM,PE:Hel1979JCP} and practically
\cite{PE:Cha1979JCP,PE:Got1976JCP} inequivalent: they are called
\emph{stiff} and \emph{rigid}. In the classical rigid model, the
constraints are assumed to be \emph{exact} and all the velocities that
are orthogonal to the hypersurface defined by them vanish. In the
classical stiff model, on the other hand, the constraints are assumed
to be \emph{approximate} and they are implemented by a steep potential
that drives the system to the constrained hypersurface. In this case,
the orthogonal velocities are activated and may act as `heat
containers'.

The conformational equilibrium of the system, according to these
models, is described by the following probability densities
\cite{PE:Ech2006JCCbNOT}:

\begin{equation}
\label{eq:correcting_terms}
\begin{array}{l@{\hspace{30pt}}l}
\mathrm{Classical\ Stiff\ Model} & \mathrm{Classical\ Rigid\ Model} \\
\hline\\[-8pt]
P_{\mathrm{s}}(q^{u}) = \frac{\displaystyle \exp
                              \big[-\beta F_{\mathrm{s}}(q^{u})\big]}
                             {\displaystyle Z^{\,\prime}_{\mathrm{s}}} & 
P_{\mathrm{r}}(q^{u}) = \frac{\displaystyle \exp
                              \big[-\beta F_{\mathrm{r}}(q^{u})\big]}
                             {\displaystyle Z^{\,\prime}_{\mathrm{r}}} \\[8pt]
F_{\mathrm{s}}(q^{u}):= V_{\Sigma}(q^{i}) -
   T \big( S_{\mathrm{s}}^{c}(q^{i}) + 
   S_{\mathrm{s}}^{\mathrm{k}}(q^{u}) \big) & 
F_{\mathrm{r}}(q^{u}):= V_{\Sigma}(q^{i}) -
   T S_{\mathrm{r}}^{\mathrm{k}}(q^{u}) \\[6pt]
\displaystyle S_{\mathrm{s}}^{\mathrm{k}}(q^{u}):=
                   \frac{R}{2}\ln\Big[\mathrm{det}\,
                   G\big(q^{u},f^{I}(q^{i})\big)\Big] & 
\displaystyle S_{\mathrm{r}}^{\mathrm{k}}(q^{u}):=
  \frac{R}{2}\ln\Big[\mathrm{det}\,g(q^{u})\Big] \\[8pt]
\displaystyle S_{\mathrm{s}}^{c}(q^{i}):=-\frac{R}{2}\ln\Big[\mathrm{det}\,
                   \mathcal{H}(q^{i})\Big] & 
\end{array}
\end{equation}

where $\beta:=1/RT$, $V_{\Sigma}$ is the potential energy in $\Sigma$
(the Potential Energy Surface (PES) frequently used in Quantum
Chemistry), and $G$, $g$ and $\mathcal{H}$ denote, respectively, the
whole-space mass-metric tensor, the reduced mass-metric tensor in
$\Sigma$ and the Hessian of the constraining part of the potential.

The different terms that correct the PES $V_{\Sigma}$ in
\eqref{eq:correcting_terms} are regarded (and denoted) as entropies
because they are linear in the temperature $T$ and come from the
averaging out of certain degrees of freedom (sometimes coordinates,
sometimes momenta). Accordingly, the effective potentials occurring in
the exponent of the equilibrium probabilities are regarded (and
denoted) as free energies.

Now, if Monte Carlo simulations in the coordinate space are to be
performed \cite{PE:Aba1994JCC,PE:Kna1993JCC} and the probability
densities that correspond to any of these two models sampled, the
correcting entropies in \eqref{eq:correcting_terms} should be included
or, otherwise, showed to be negligible.

On the other hand, if rigid Molecular Dynamics simulations are
performed with the intention of sampling from the \emph{stiff}
equilibrium probability $P_{\mathrm{s}}$
\cite{PE:Pas2002JCP,PE:He1998JCPb,PE:Fix1978JCP}, then, the so-called
\emph{Fixman's compensating potential} \cite{PE:Fix1974PNAS},

\begin{equation}
\label{eq:VF}
V_{\mathrm{F}}(q^{u}) := T S_{\mathrm{r}}^{\mathrm{k}}(q^{u}) -
 T S_{\mathrm{s}}^{c}(q^{i}) - 
 T S_{\mathrm{s}}^{\mathrm{k}}(q^{u}) = 
  \frac{RT}{2}\ln\Bigg [\frac{\mathrm{det}\,G(q^{u})}{\mathrm{det}\,
  \mathcal{H}(q^{i})\, \mathrm{det}\,g(q^{u})}\Bigg] \ ,
\end{equation}

must be added to the PES $V_{\Sigma}$.

The conformational dependence of most of the determinants appearing in
\eqref{eq:correcting_terms} and \eqref{eq:VF} is frequently assumed to
be negligible in the literature and they are consequently dropped from
the calculations
\cite{PE:All2005BOOK,PE:Pat2004JCP,PE:Fre2002BOOK,PE:Go1976MM}. Also,
subtly entangled to the assumptions underlying these simplifications,
a second type of approximation is made that consists of assuming that
the equilibrium values of the hard coordinates do not depend on the
soft coordinates
\cite{PE:All2005BOOK,PE:Mor2004ACP,PE:Pat2004JCP,PE:Fre2002BOOK}.
This has been argued to be only approximate even in the case of
classical force fields
\cite{PE:Che2005JCC,PE:Hes2002JCP,PE:Zho2000JCP}.

In \cite{PE:Ech2006JCCbNOT}, we have eliminated all simplifying
assumptions and measured the conformational dependence on the
Ramachandran angles $\phi$ and $\psi$ (the soft coordinates) of
\emph{all correcting terms} and of the Fixman's compensating potential
in the model dipeptide HCO-{\small L}-Ala-NH$_2$. The potential energy
function used was the effective Born-Oppenheimer potential for the
nuclei (see \eqref{eq:Schroedinger}) derived from ab initio quantum
mechanical calculations including electron correlation at the
\mbox{MP2/6-31++G(d,p)} level of the theory.

\begin{figure}
  \includegraphics[height=.3\textheight]{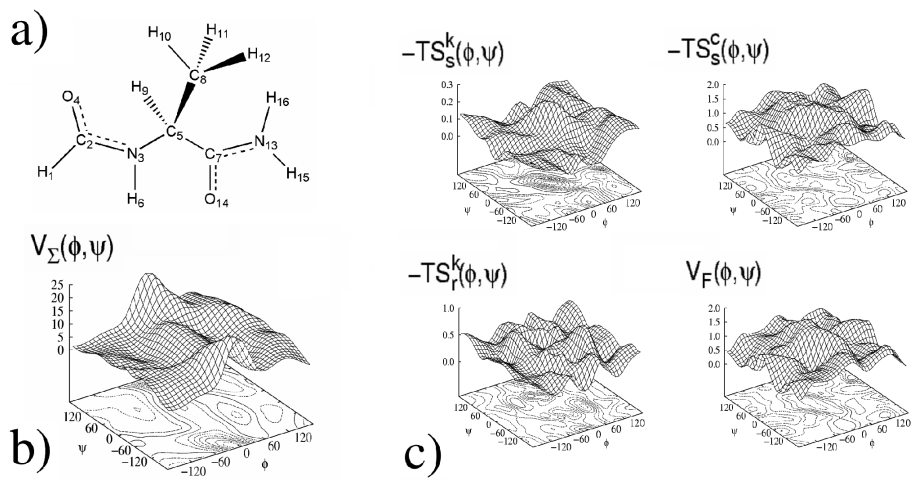}
  \caption{{\bf a)} Model dipeptide HCO-L-Ala-NH$_2$ numbered
  according to the SASMIC \cite{PE:Ech2006JCCaNOT} scheme. \mbox{{\bf
  b)} Potential} Energy Surface. {\bf c)} Conformational dependence of
  the correcting terms. All energies are given in kcal/mol.}
\end{figure}

\begin{table}[!ht]
\begin{tabular}{cccrrrr}
\hline
  \tablehead{1}{c}{}{Corr.\tablenote{Correcting term whose importance is measured in the corresponding row}}
& \tablehead{1}{c}{}{${V_{1}}$\tablenote{`Correct' potential energy; the one containing the correcting term}}
& \tablehead{1}{c}{}{${V_{2}}$\tablenote{`Approximate' potential energy; the one lacking the correcting term}}
& \tablehead{1}{c}{}{${d_{12}}$\tablenote{Statistical distance between $V_{1}$ and $V_{2}$ (see \cite{PE:Alo2006JCC})}}
& \tablehead{1}{c}{}{${N_{\mathrm{res}}}$\tablenote{Number of residues in a polypeptide potential up to which the correcting term may be omitted}}
& \tablehead{1}{c}{}{${b_{12}}$\tablenote{Slope of the linear rescaling between $V_{1}$ and $V_{2}$}}
& \tablehead{1}{c}{}{${r_{12}}$\tablenote{Pearson's correlation coefficient}}
\\
\hline
$- TS_{\mathrm{s}}^{\mathrm{k}} - TS_{\mathrm{s}}^{\mathrm{c}}$ &
 $F_{\mathrm{s}}$ & $V_{\Sigma}$ &
 0.74 $RT$ & 1.82 & 0.98 & 0.9967 \\
$- TS_{\mathrm{s}}^{\mathrm{c}}$ &
 $F_{\mathrm{s}}$ & $V_{\Sigma} - TS_{\mathrm{s}}^{\mathrm{k}}$ &
 0.74 $RT$ & 1.83 & 0.98 & 0.9967 \\
$- TS_{\mathrm{s}}^{\mathrm{k}}$ &
 $F_{\mathrm{s}}$ & $V_{\Sigma} - TS_{\mathrm{s}}^{\mathrm{c}}$ &
 0.11 $RT$ & 80.45 & 1.00 & 0.9999 \\[6pt]
$- TS_{\mathrm{r}}^{\mathrm{k}}$ &
 $F_{\mathrm{r}}$ & $V_{\Sigma}$ &
 0.29 $RT$ & 11.62 & 1.01 & 0.9995 \\[6pt]
$V_{\mathrm{F}}$ & $F_{\mathrm{s}}$ & $F_{\mathrm{r}}$ &
 0.67 $RT$ & 2.24 & 0.97 & 0.9972 \\
\hline
\end{tabular}
\caption{\label{tab:distances}Quantitative assessment of the
importance of the different correcting terms involved in the study of
the constrained equilibrium of the protected dipeptide
HCO-L-Ala-NH$_2$ (see \cite{PE:Ech2006JCCbNOT}).}
\end{table}

In table \ref{tab:distances}, the main results of our work are
presented. The importance of all the correcting terms is assessed by
comparing (with the statistical distance $d_{12}$ described in the
previous section) the effective potential $V_{1}$, containing the term,
with the approximate one $V_{2}$, lacking it. Moreover, if one assumes
that the effective energies compared will be used to construct a
polypeptide potential, the number $N_{\mathrm{res}}$ of
residues up to which one may go keeping the distance between the two
approximations of the the $N$-residue potential below $RT$ is (see
eq.~(23) in \cite{PE:Alo2006JCC}):

\begin{equation}
\label{eq:Nres}
N_{\mathrm{res}}=\left ( \frac{RT}{d_{12}} \right )^{2} \ .
\end{equation}

In the table, one can see that, in the stiff model, the
Hessian-related correcting term should be included in Monte Carlo
simulations for peptides as short as two residues, while the one that
depends on $G$ may be neglected up to chains which are $\sim$ 80
residues long. The only correcting term occurring in the rigid model,
in turn, may be dropped up to $\sim$ 12 residues. Finally, the Fixman
potential, containing all determinants, should be included in MD rigid
simulations of peptides with more than two residues\footnote{One
should note that the distance between the PES $V_{\Sigma}$
at \mbox{MP2/6-31++G(d,p)} and the one computed at
\mbox{HF/6-31++G(d,p)} is $d_{12} \simeq$ 1.2 $RT$. A value
slightly larger but of the order of the ones obtained when
the most important correcting terms are dropped.}.

These results are related to a working set of conformations consisting
of 144 points regularly distributed in the whole Ramanchandran
space. In a second part of the work, we have repeated all the
comparisons for a working set consisting of six secondary structure
elements. The results suggest that, if one is interested only in this
energetically lower region, the distances $d_{12}$ are roughly divided
by two and, accordingly, the values of $N_{\mathrm{res}}$ are four
times larger.

We have also repeated the calculations, with the same basis set
(6-31++G(d,p)) and at the Hartree-Fock level of the theory in order to
investigate if this less demanding method without electron correlation
may be used in further studies. We have found that, indeed, this can
be done, obtaining very similar results at a tenth of the
computational effort.

As far as we are aware, this is \emph{the first time} that this type
of study is performed in a relevant biomolecule with a realistic
potential energy function.

\begin{theacknowledgments}
We thank F. Falceto, V. Laliena, F. Plo and D. Zueco for illuminating
discussions. The numerical calculations have been performed at the
BIFI computing center. We thank I. Campos, for the invaluable CPU time
and the efficiency at solving problems.

This work has been supported by the Arag\'on Government
(``Biocomputaci\'on y F\'{\i}sica de Sistemas Complejos'' group) and
by the research grants MEC (Spain) \mbox{FIS2004-05073} and
\mbox{FPA2003-02948}, and MCYT (Spain) \mbox{BFM2003-08532}. P.
Echenique and I. Calvo are supported by MEC (Spain) FPU grants.
\end{theacknowledgments}

\bibliography{pablo_echenique}

\end{document}